\documentclass[prl,showpacs,footinbib,twocolumn,tightenlines,final,lengthcheck,letterpaper,amsmath,english]{revtex4}

\usepackage[T1]{fontenc}
\usepackage[utf8]{inputenc}
\usepackage[pdftex]{color,graphicx}
\usepackage[pdftex]{hyperref}

\hypersetup{colorlinks=false}
\usepackage[english]{babel}
\usepackage{dcolumn}

\newcommand{\ket}{\rangle}
\newcommand{\bra}{\langle}
\newcommand{\spc}{\big|}
\newcommand{\R}{\mathrm}
\newcommand{\reff}[1]{(\ref{#1})}

\newcommand{\Sr}[1]{${}^{#1}\mathrm{Sr}$}
\newcommand{\s}[1]{{\hspace{#1pt}}}

\makeatletter \renewcommand{\fnum@figure} {FIG. \thefigure} \long\def\@makecaption#1#2{ \par
  \vskip\abovecaptionskip
  \begingroup
   \small\rmfamily
   \sbox\@tempboxa{    \let\\\heading@cr
    #1. #2   }   \@ifdim{\wd\@tempboxa >\hsize}{    \begingroup
     \samepage
     \flushing
     \let\footnote\@footnotemark@gobble
     #1. #2\par\vspace{-3mm}
    \endgroup
   }{     \global \@minipagefalse
     \hb@xt@\hsize{\hfil\unhbox\@tempboxa\hfil}   }  \endgroup
  \vskip\belowcaptionskip
}

\def\bib@device#1#2{} 
\makeatother

\begin{document}

\title{Coherent transport of atomic wave packets in amplitude-modulated 
vertical optical lattices}

\author{A. Alberti} \author{G. Ferrari} \author{V. V. Ivanov} 
\author{M. L. Chiofalo\footnote{Permanent address: Department of Mathematics and INFN, 
University of Pisa, Largo B. Pontecorvo 5, 56127 Pisa, Italy}} 
\author{G. M. Tino} \email{guglielmo.tino@fi.infn.it} 
\affiliation{Dipartimento di 
Fisica and LENS - Universit\`a di Firenze, CNR-INFM,\\
INFN - Sezione di 
Firenze, via Sansone 1, 50019 Sesto Fiorentino, Italy}

\begin{abstract} 
We report on the realization of
dynamical control of transport for ultra-cold \Sr{88} atoms loaded in an 
accelerated and amplitude-modulated 1D optical lattice. We 
tailor the energy dispersion of traveling wave packets and
reversibly switch between Wannier-Stark localization 
and driven transport based on coherent tunneling. Within a Loschmidt-echo scheme where the atomic group velocities are reversed at once, 
we demonstrate a novel mirror for matter waves working
independently of the momentum state and discuss possible applications 
to force measurements at micrometric scales. 
\end{abstract}

\pacs{03.75.Lm, 37.10.Jk, 03.65.Yz, 04.80.-y}

\keywords{driven optical lattice, atom mirror, Bloch oscillations, gravity}

\maketitle 

\newlength{\oldtextheight}\oldtextheight=\textheight 

Atoms trapped in optical lattice potentials in the  
absence of defects and phonon excitations, are extremely versatile systems 
for quantum applications~\cite{Bloch:2008p1173} including
transport~\cite{Raizen:1997p753} and strongly correlated phases 
\cite{Greiner:2002p709}.
In addition, control of atomic interactions may keep the dynamics coherent over 
seconds~\cite{Ferrari:2006p532}, and has 
led to atomic optical clocks with extraordinary 
performances~\cite{Blatt:2008p140801}, determination
of forces at micrometer resolution 
\cite{Anderson:1998p517,Roati:2004p1040,Ferrari:2006p532,Wolf:2007p986}, and quantum 
information processing~\cite{Bloch:2008p128}.

Tailoring and control of the coherent transport behavior is an essential tool for these 
applications~\cite{Grifoni:1998p842}. Dynamical mechanisms, such as 
phase or amplitude lattice modulations, can be designed to drive 
trasport behavior over desired time and length scales. 
Phase modulation, obtained by a periodic spatial displacement of the optical 
lattice, has been
earlier exploited to investigate quantum chaos~\cite{Hensinger:2001p52} and,  
under the effect of constant forces, to observe Wannier-Stark 
resonances~\cite{Wilkinson:1996p710},   
photon-assisted resonant tunneling~\cite{Ivanov:2008p1}, 
quantum transport over macroscopic distances~\cite{Alberti09:Nature}, 
and dynamical control 
of Mott insulator transitions~\cite{Zenesini:2009p1406}. 
Amplitude driving has been used so far as a spectroscopic
tool for interband excitations~\cite{Denschlag:2002p3095}, and 
 to characterize the Mott insulator 
regime~\cite{Stoferle:2004p130403}. Analogous techniques are produced by acoustoelectric means 
in semiconductors nanostructures in the form 
of surface acoustic waves~\cite{Gell:2008APLp81115}, but in this case the relevant spatial scale is much longer than the 
underlying static lattice.

In this Letter we report on dynamical control of coherent transport  
of quantum wave packets which is realized through amplitude modulation of vertical optical lattices.
The control of transport allows to reverse all the atomic velocities at once, as in an atom mirror, assessing an effect similar to
Loschmidt echos~\cite{Cucchietti:2003p210403}, and it provides a new tool for precise force measurements with micrometric spatial resolution. While extensive theoretical studies on phase-modulation of optical lattices are 
available~\cite{Dunlap:1986p841,Thommen:2002p698,Eckardt:2005p769}, the  
physics underlying the control of transport through amplitude modulation  
is to our knowledge largely unexplored. 
 
\begin{figure}[htbp] \begin{center} 
\includegraphics[width=0.47\textwidth]{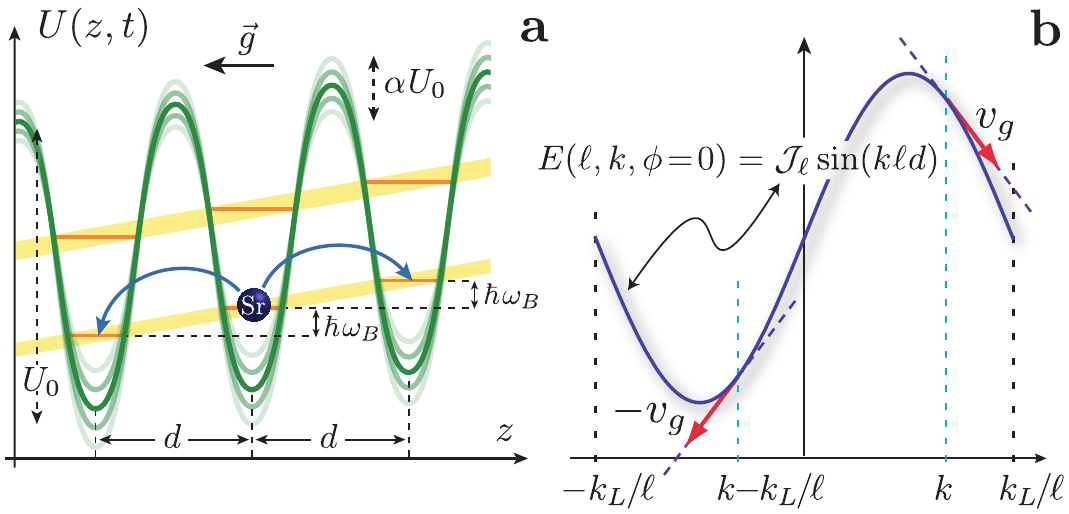} 
\caption{\label{fig:figure1} 
Sketch of the transport mechanism in the amplitude-modulated (AM) lattice.
(a) In the real space: Wannier-Stark localized states populate the tilted lattice of energy potential $U(z,t)$. In the absence of AM, $\alpha=0$,  the intersite tunneling is suppressed. Coherent tunneling between sites $\ell\s{0.5}d$ apart
is enabled when AM is switched on at $\omega_M=\ell\s{0.3}\omega_B$. 
(b) In the momentum space: wave packets behave as if they were moving in an effective sinusoidal energy band, 
whose width, periodicity and phase can be tailored by means of the AM parameters. 
Because of the sinusoidal shape, the group velocity $v_g$ is reversed when $k\rightarrow k-k_L/\ell$.} 
\end{center} 
\end{figure}

In the experiment, a 1D optical lattice for \Sr{88} atoms is originated by the interference pattern of two 
vertical counter-propagating laser beams with wavelength $\lambda_L$, so that  
atoms effectively see a periodic potential 
with a period $d=\lambda_L/2$ and a depth
dictated by the laser intensity, which we modulate in time. Our
system is then described by the 1D time-dependent single-particle Hamiltonian: 
\begin{equation} 
\label{eq:hamiltonian} 
\mathcal{H}(z,p,t) =\frac{p^2}{2m}
-U(z)\left[1+\alpha f(t)\right]+mgz 
\end{equation} 
where $U(z)=U_0\cos(2k_L z)/2$ and 
$f(t)\equiv\sin\s{-2}\big[\omega_M(t-t_0)-\phi\big]$. 
As in Fig.~\hyperref[fig:figure1]{\ref*{fig:figure1}a}, 
$U_0$ is the lattice depth, $k_L=2\pi/\lambda_L$ is the laser wave 
vector, $m$ is the atomic mass, $g$ is the gravity 
acceleration along the lattice direction, $0<\alpha<1$ is the modulation 
amplitude expressed in units of $U_0$, $\omega_M=\ell\s{0.8}\omega_B$ is the $\ell$--th harmonic of the 
Bloch frequency $\omega_B=m\s{0.5}g\s{0.5}d/\hbar$, and $t_0$ the 
time when the modulation is switched on with initial phase $\phi$. When $\alpha=0$, 
the resulting static Hamiltonian $\mathcal{H}_0$ is diagonalized by the so-called
Wannier-Stark states $\spc n \ket$ centered at the 
$n$--th lattice site and separated in energy by quanta of $\hbar\s{0.3}\omega_B$.
In these conditions, 
wave packet tunneling is frozen due to 
Wannier-Stark localization and 
Bloch oscillations occur~\cite{Raizen:1997p753}, where 
the quasi-momentum spans the Brillouin zone $[-k_L,k_L]$ 
with period $\tau_B=2\pi/\omega_B$. 

When the modulation is on at frequency $\ell\s{0.5}\omega_B$ a coherent resonant tunneling is established between sites spatially separated by $\ell\s{0.5}d$ \cite{Ivanov:2008p1}, as sketched in Fig.~\hyperref[fig:figure1]{\ref*{fig:figure1}a}. In the rotating wave approximation, the application of 
the unitary transformation $\mathcal{U}=\exp(-i\mathcal{H}_0t)$ simplifies
the Hamiltonian of Eq.~\reff{eq:hamiltonian} into 
$\mathcal{H}'\s{0}=\s{-7}\sum_{n=-\infty}^{\infty}\s{-4} \left[\s{-1}i 
({\mathcal{J}_\ell}/{2})e^{i\phi}\spc n\s{-1}+\s{-1}\ell\s{0.5} \ket \bra 
n\spc\s{-1}+\s{-1}\R{h.c.}\s{-1}\right]$ with tunneling  rates $\mathcal{J}_\ell=-\alpha U_0\bra n\s{-1}+\s{-1}\ell\s{0.5}\spc\s{-1}\cos(2k_Lx)\spc n\ket/2$ \cite{resonant_terms}.
This site-to-site tunneling can be equivalently viewed as 
a two-photon stimulated Raman process induced by the electric field of the lattice laser which acquires sidebands displaced by $\ell\s{0.3}\omega_B$~\cite{Denschlag:2002p3095}, and it offers an alternative approach to \cite{Wolf:2007p986}. In this pictures the $\mathcal{J}_\ell$ coefficients play the role of Frank-Condon factors~\cite{Deb:2007p4131}.
In the momentum space, the atomic wave packets behave as if they were  
moving with an effective energy dispersion:
\begin{equation} 
\label{eq:effective_dispersion_law} 
E(\ell,k,\phi)=\mathcal{J}_\ell\sin(k\s{0.5}\ell\s{0.5} d-\phi)
\end{equation} 
as depicted in Fig.~\hyperref[fig:figure1]{\ref*{fig:figure1}b}, where 
each of the three parameters $\mathcal{J}_\ell$, $\ell$ and $\phi$ can be accurately tuned. \textheight=\oldtextheight This tunable sinusoidal energy profile has deep consequences 
for the trasport behavior. First, the size of the 
Brillouin zone is effectively reduced by a factor $\ell$. Contrary 
to the subwavelength lattice studied in~\cite{Yi:2008p73015,Salger:2009p11605}, 
this is equivalent to establish an effective super-lattice
period $\ell\s{0.5} d$ where the modulation at the $\ell$--harmonics of 
$\omega_B$ controls the addressing of the $\ell$--th site. 
Second, both $\phi$ and the energy bandwidth $\mathcal{J}_\ell$ can be 
tuned, the latter being linear in the full range of 
modulation amplitudes $0<\alpha<1$. 
Finally, the group velocity $v_g(\ell,k,\phi)=\partial E(\ell,k,\phi)/\partial 
(\hbar k)=\ell \s{0.5}d\s{0.5}\mathcal{J}_\ell\cos[k\s{0.5}\ell\s{0.5} d-\phi]/\hbar$ 
reverses its sign whenever $k\rightarrow 
k+(2\s{0.3}n+1)\,k_L/\ell$ for any integer $n$, as shown in Fig.~\hyperref[fig:figure1]{\ref*{fig:figure1}b}. This recalls the Loschmidt-echo scheme in~\cite{Cucchietti:2003p210403}, where it is shown that the wave vector mapping is equivalent to reversing the band curvature $\mathcal{J}_\ell\rightarrow -\mathcal{J}_\ell$.     
This reversal can be a powerful tool to study
decoherence processes and fidelity in 
quantum many-body systems~\cite{Zurek:2003p715}.
We thus proceed to illustrate the experiment where this velocity reversal is 
performed.

The source of ultra-cold atoms has been described elsewhere 
\cite{Ferrari:2006p532}. Shortly, we start by trapping and cooling about 
$2 \times 10^7$ \Sr{88} atoms at $3\;\R{mK}$ in a magneto-optical trap 
(MOT) operating on the $^1\s{-1}S_0-^1\s{-2.5}P_1$ resonance transition 
at $461\;\R{nm}$. The temperature is further reduced by a second cooling 
stage in a red MOT operating on the $^1\s{-1}S_0-^3\s{-2.5}P_1$ 
intercombination transition at $689\;\R{nm}$. Finally we obtain $\sim 
1\times 10^6$ atoms at $1\;\R{\mu K}$. We load the atoms in the optical 
lattice which is switched on adiabatically in $80\;\R{\mu s}$. The 
atomic sample arranges itself in a disk-shaped geometry with a vertical 
RMS size of about $\sigma_0=30\;\R{\mu m}$. Atomic interactions here are negligible because 
of the tiny scattering length in the ground state $a=-1.4\,a_0$ \cite{MartinezdeEscobar:2008p1192}. The lattice potential is originated 
by a single-mode frequency-doubled $\textrm{Nd:YVO}_4$ laser ($\lambda_L 
= 532\;\R{nm}$) delivering up to $1\;\R{W}$ on the atoms with a beam 
waist of $250\;\R{\mu m}$. The beam is vertically aligned and 
retro-reflected by a mirror. The resulting Bloch frequency is 
$\omega_B=2\pi\times574.3\;\R{s^{-1}}$. The corresponding photon recoil 
energy is $E_R = h^2/(2m \lambda_L^2)=2\pi\times8000\;\R{s}^{-1}$, and  
lattice depth ranges from $5$ to $20\;E_R$, when 
the energy gap at $k_L$ 
is $E_G\gtrsim3\,E_R\gg\hbar\omega_B$ and the bandwidth is always smaller than $10^{-1}E_G$. Given these conditions, 
Landau-Zener tunneling is negligible. By controlling the 
radio-frequency power of an acousto-optical modulator we stabilize and 
modulate the laser intensity in order to reproduce the time-dependent 
Hamiltonian in Eq.~\ref{eq:hamiltonian}. The readout is performed by 
measuring \emph{in situ} the spatial atomic distribution using resonant 
absorption imaging.

To characterize the tunneling rates $\mathcal{J}_\ell$ we proceed as follows. The atoms are initially loaded in 
single lattice sites, their quasi momenta being spread 
over the whole Brillouin zone. In fact, the de Broglie wavelength results $\lambda_\R{dB}\sim 200\,\R{nm}<d$ at temperature $T\sim 1\,\R{\mu K}$.  Applying the amplitude modulation for a given time results 
into a spatial broadening of the atomic wave packets. 
The speed of this broadening corresponds to the average over $k$ of the group velocities $\langle{}v_g^2\rangle{}^{1/2}=\ell \s{0.5}d\s{0.5}\mathcal{J}_\ell/(\sqrt{2}\hbar)$ which, thus, provides a measurement of $\mathcal{J}_\ell$.
We measure the speed of broadening varying the modulation amplitude $0<\alpha<1$, and find that the $\mathcal{J}_\ell$ depend linearly on it. This is a unique feature of amplitude modulation which provides a clean tunability of the $\mathcal{J}_\ell$ in a wide range. As a function of $U_0$ and $\ell$, we observe that the tunneling rates are well fitted by
$|\mathcal{J}_\ell(U_0/E_R)|\approx|\mathcal{J}_1(U_0/E_R)|\exp[-\beta_1\s{1.7}(\ell-1)\s{0.5}U_0/E_R]$
with $\beta_1=0.35$ and $|\mathcal{J}_1(U_0/E_R)|/\hbar \approx 2500 \times (  \alpha U_0/E_R ) \exp(-\beta_2\s{1.7}\s{0.5}U_0/E_R)\, \R{s}^{-1}$ with $\beta_2=0.25$. This agrees with the  
expression which can be calculated employing the Wannier-Stark basis.
\begin{figure}[t] \begin{center} 
\includegraphics[width=0.485\textwidth]{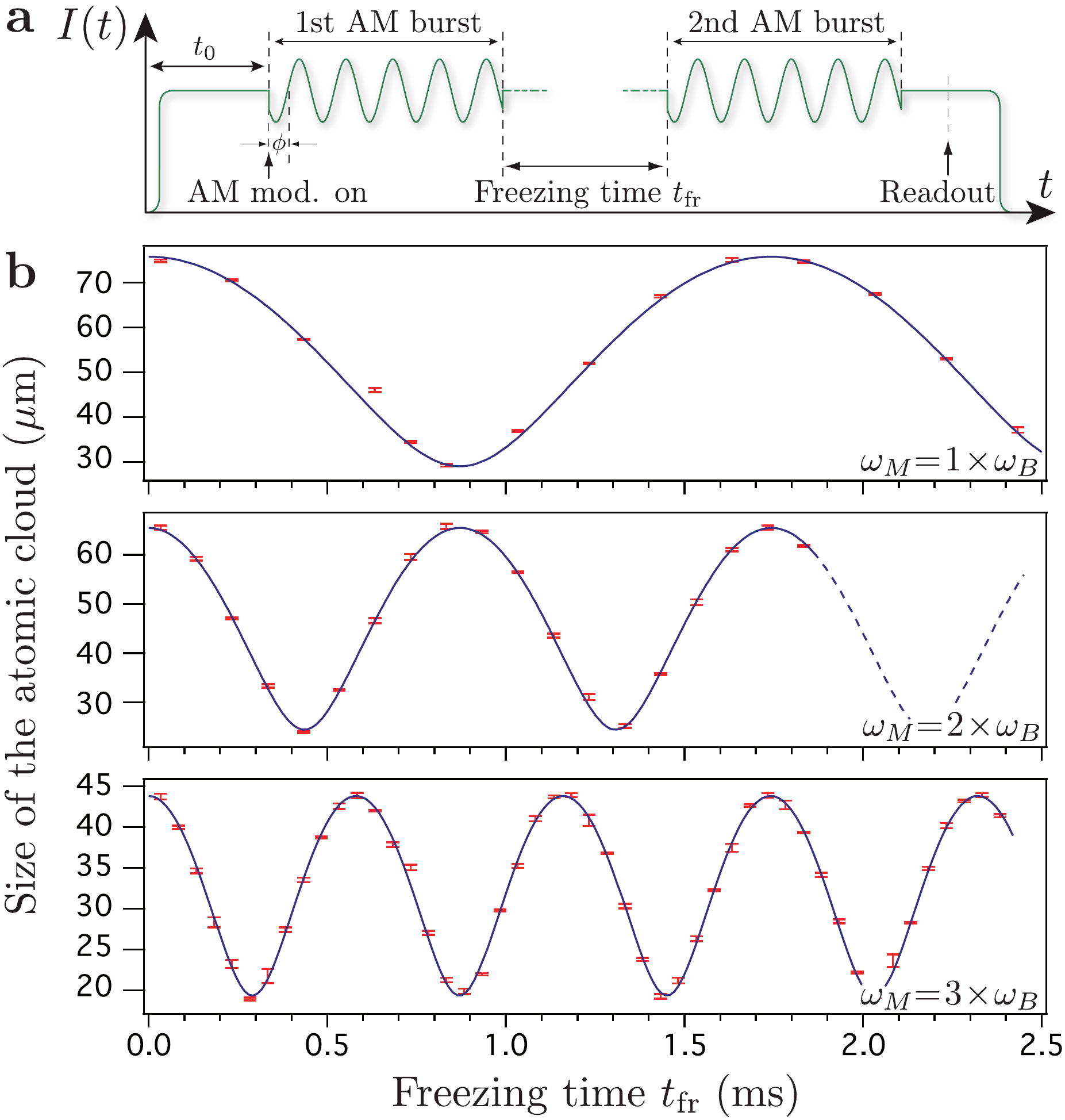} 
\caption{\label{fig:figure2}
Realization of Loschmidt echos 
(a) The applied sequence: two identical AM bursts of the optical lattice intensity $I(t)$ 
at $\ell\omega_B$ enable the resonant tunneling which couples lattice sites separated by $\ell d$. The bursts are 
separated by a \emph{freezing time} $t_\R{fr}$, during which the tunneling is disabled and
Bloch oscillations of the quasi momentum set in. (b) The corresponding cloud's RMS size \emph{vs.}~$t_\R{fr}$ has periodicity $2\pi/(\ell\s{0.4}\omega_B)$. From top to bottom, 
AM at $\omega_M=\ell\omega_B$ induces 
resonant coupling with up to $\ell=3$ neighboring sites. Inversion 
of the group velocities $v_g$ occurs at the minima of the RMS size.}
\end{center} 
\end{figure}

In order to demonstrate the reversal of the group 
velocities, we employ the sequence in Fig.~\hyperref[fig:figure2]{\ref*{fig:figure2}a}, using a Loschmidt-echo scheme analogous to the more familiar spin echo in other systems \cite{Cucchietti:2003p210403}. 
Two identical AM bursts at $\ell\omega_B$ are applied 
lasting $287\,\tau_B\sim 500\;\R{ms}$ each, where $\tau_B$ has been preliminarily measured with 
$1\;\R{ppm}$ sensitivity~\cite{Ivanov:2008p1}. For each case $\ell=1,2,3$, 
$U_0=11.2,\,6.6,\,6.3\;E_R$ and $\alpha=0.23,0.47,0.84$ are chosen 
to keep comparable tunneling rates. The two bursts serve to enable 
the coherent tunneling and are separated by a 
variable \emph{freezing time} $t_\R{fr}$ (with $t_0=0$ and $\phi=0$), during which 
AM is off and tunneling is disabled. During the \emph{freezing time}, Bloch oscillations occur in the reduced Brillouin zone $[-k_L/\ell,k_L/\ell]$
which is spanned by the atomic wave vector $k$ with an effective Bloch period $\tau_\ell=\tau_B/\ell$. 
The magnitude and sign of $v_g(\ell,k,\phi)$ change while $k$ spans the Brillouin zone, so that 
the atomic cloud is expected to expand or shrink depending 
on the value of $v_g$ reached at the start of the second burst. 
Fig.~\hyperref[fig:figure2]{\ref*{fig:figure2}b} shows that this 
is indeed the case, resulting in a periodic signal with period $\tau_\ell$. 
The group velocities $v_g(\ell,k,\phi)$ are fully reversed independently of the $k$ values 
at each $\tau_\ell/2$, that is at the minima occurring at $t_\R{fr}=\tau_\ell/2|2n+1|$ ($n$ integer). 
The Gaussian convolution of the single atom response with the initial cloud's distribution suggests that the expression of the RMS size is
$\sigma(t_\R{fr})=\sqrt{\sigma_0^2+\sigma_1^2\cos(\pi\, t_\R{fr}/\tau_\ell)^2}$, with $\sigma_0$ the initial size.
This is
reproduced by the solid curves, best agreeing with the measured points. The fit to each data 
set with different $\ell$ yields 
$\tau_\ell=\tau_B/\ell$, and $\sigma_1$ turns out to match the broadening which would occur after one single burst twice as long.
The oscillations persist undamped over several 
seconds, indicating a high fidelity in recovering
the quantum state after $v_g$ reversal~\cite{Cucchietti:2003p210403}. 
\begin{figure}[hb] \begin{center} 
\includegraphics[width=0.485\textwidth]{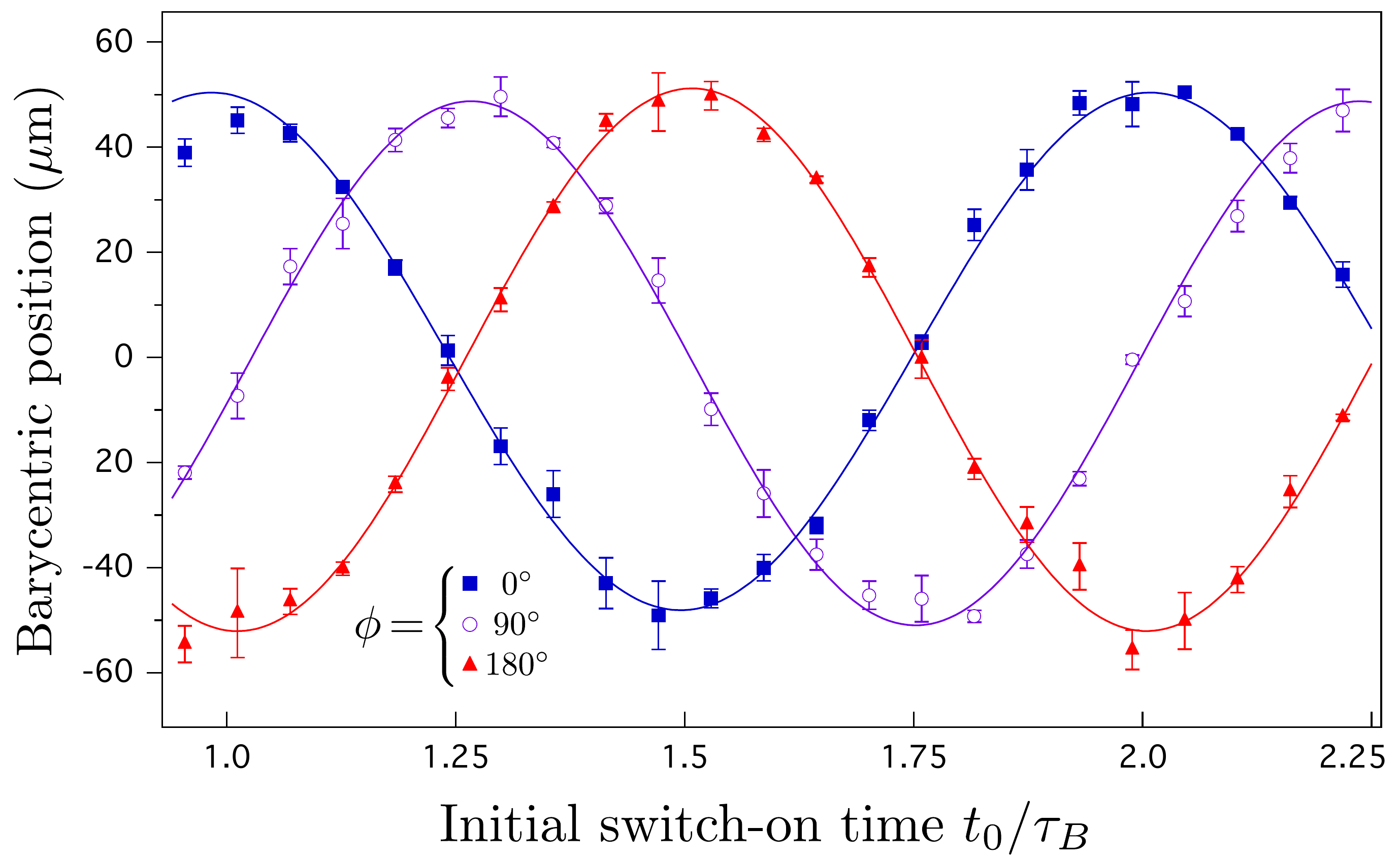} 
\caption{\label{fig:figure3}
Cloud's barycentric position after a single AM burst \emph{vs.} swtich-on time of the AM modulation, $t_0$, at
$\omega_M=\omega_B$. By varying the modulation phase $\phi$ of the driving we shift the sinusoidal dispersion law
by the same amount, \emph{e.g.}~at $\phi=90^\circ$ from sine to cosine.
} 
\end{center} 
\end{figure}

\begin{figure}[t] \begin{center} 
\includegraphics[width=0.485\textwidth]{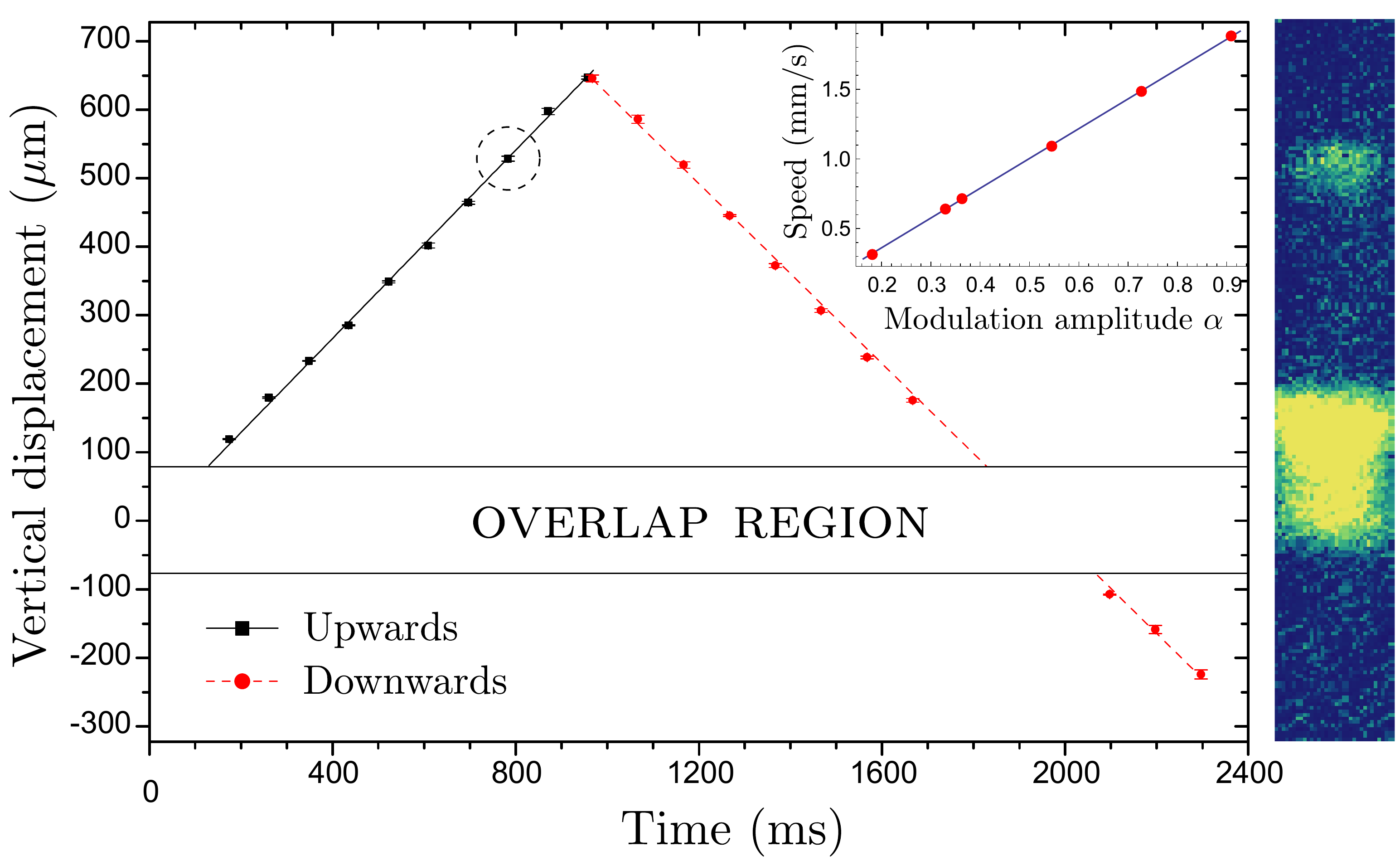} 
\caption{\label{fig:figure4} Atom mirror for traveling matter waves. 
Atoms loaded in the 2nd band 
have sub-recoil momentum dispersion. 
Wave packets with initial $k=0$ 
travel upwards with minor dispersion (solid line). After reversing the group velocity $v_g$, 
atoms move downwards with opposite speed (dashed line). Inset: linearity of the tunneling \emph{vs} the modulation amplitude $\alpha$. Right 
panel: 2D density profile of the cloud corresponding to the data point, circled in figure, at
$t=783\;\R{ms}$. Since the atoms in the 2nd band have an higher $v_g$, the two clouds separate each other away from the origin.}
\end{center}
\end{figure}

In order to observe the mirror effect for traveling matter waves with defined momentum, 
we need to prepare the atoms with a momentum dispersion narrower than the 
Brillouin zone. We accomplish this by 
increasing $U_0=14\;E_R$ sufficiently to trap a 
fraction of atoms also in the second band, where the momentum distribution shrinks
since here the thermal momentum distribution is steeper. In 
addition, we purposely favor one of the two directions by letting the 
atoms freely fall for about $200\;\R{\mu s}$ between the release from 
the red MOT and the lattice switch on. We can accurately control 
the value of $k$ within the Brillouin zone by following the Bloch oscillations in the static lattice for a 
time $t_0$ before activating the amplitude modulation. During the modulation the cloud's position is expected to move with a velocity $v_g(\ell,k,\phi)$ with $k=k_L\s{1}t_0/\tau_B$, as it is shown in Fig.~\ref{fig:figure3} . At 
the flex points of $E(\ell,k,\phi)$, like at $k=0$ for $\phi=0$, 
the motion becomes dispersionless and 
reaches the largest $v_g$, and consequently the largest displacement. In this case, Fig.~\ref{fig:figure4} shows the displacement of the atoms in the second band which move upwards with a 
$v_g$ of $0.64\;\R{mm}/\R{s}$ when $\alpha=0.33$ and $\ell=1$. We track this motion for $500$ 
cycles on a distance of about $0.5\;\R{mm}$. A residual broadening still 
occurs because of finite spread of the initial momentum distribution, but 
it is limited to 1/10 of the total displacement. 
This can be largely reduced by initially preparing a narrower momentum distribution, as with Bragg or Raman velocity selective stages, sympathetic cooling or BECs.
In order to 
reverse the wave-packets' motion we subsequently apply the scheme 
of Fig.~\hyperref[fig:figure2]{\ref*{fig:figure2}a} by 
choosing $t_\R{fr}=\tau_B/2$.  During the second burst the atoms move 
downwards with the same speed but opposite direction.
As expected, the atom mirror reverses the direction of motion as well as the broadening of the atomic cloud, 
making the traveling wave packets recover their initial size after $500$ 
cycles of the second burst. 

Tailoring the transport with AM in a lattice could provide an excellent tool in the study of potentials at short distances from a surface, e.g. the Casimir-Polder interaction, since it allows to measure forces with micrometric spatial resolution and high sensitivity. For instance, we report a measurement of $g$ with relative sensitivity \emph{e.g.}~$\Delta g/g =5\times 10^{-7}$, which is obtained following the
oscillations in Fig.~\hyperref[fig:figure2]{\ref*{fig:figure2}b} at $\omega_M=3\s{1}\omega_B$ for 7 seconds. The amplitude modulation method can also be employed in a single pulse scheme as in~\cite{Ivanov:2008p1} for phase modulation, with the significant advantage that now it can be implemented in the vicinity of a reflective substrate since an independent control of the two lattice beams is not required. In addition, compared to other methods based on the study of Bloch oscillations in static lattices \cite{Roati:2004p1040,Sorrentino:2009p013409}, our method has the advantage of being insensitive to the temperature of the atomic sample, and being suited to measure potential profiles even with strong curvature.

In conclusion, we presented a novel method to tailor the transport of cold atoms in optical lattices. By modulating the intensity of the lattice we are able to modify the fundamental properties governing the transport such as the dimension of the Brillouin zone and the width of the sinusoidal energy band. This allows us to realize a Loschmidt-echo scheme, offering a new tool for the study of decoherence phenomena as well as for applications in atom optics. Force measurements with sub-ppm sensitivity and micrometric spatial resolution were obtained, suggesting additional applications in the field of force measurements at short distances from a surface.

\begin{acknowledgments} We thank M. Artoni for
critical reading, M. Schioppo for experimental 
assistance, V. Piazza for helpful discussions, and R. Ballerini, M. De Pas, M. Giuntini, A. Hajeb, A. 
Montori for technical assistance. This work was supported by LENS, INFN, 
EU (under contract RII3-CT-2003 506350 and FINAQS), ASI. 
\end{acknowledgments}

\bibliographystyle{apsrev}

\begin{thebibliography}{39}
\bibitem[resontant()]{resonant_terms}
Since only resonant terms contribute to the tunneling,
  off-resonant interband transitions can be safely neglected, as well as
  Landau-Zener tunneling which is strongly suppressed in our experimental
  conditions.
\end{thebibliography}

\end{document}